\documentclass[hyperref,UTF8]{article}
\usepackage[colorlinks,linkcolor=blue,anchorcolor=yellow,citecolor=blue,urlcolor=blue]{hyperref}
\usepackage{geometry}
\usepackage{subfigure}
\usepackage{xcolor}
\usepackage{colortbl,dcolumn}
\usepackage{graphicx}
\usepackage{indentfirst}
\usepackage{listing}
\usepackage{fancyhdr}
\usepackage{amsmath}
\usepackage{amssymb}
\usepackage{multicol}
\usepackage{caption}
\usepackage{empheq}
\usepackage{tcolorbox}
\usepackage{authblk}
\usepackage{soul}

\title{Alias Sampling Effect on the Calculation of MHD Mode Number in Fusion Plasma}
\author{Nan CHU}
\affil{Institute of Plasma Physics, Chinese Academy of Sciences, Hefei 230031, China}

\definecolor{myblue}{rgb}{0.8, 0.8, 1}

\begin{document}
\maketitle{\large}
\thispagestyle{empty}

\begin{abstract}
Detailed alias frequency formula and the effect of alias sampling on the calculation of MHD mode number are derived. It is discovered that the absolute MHD mode number/structure does not change under alias sampling. This discovery can help us to determine the structure of the high frequency MHD mode with low frequency sampled diagnostics even when the Nyquist-Shannon sampling theorem is no longer valid.
\end{abstract}

\section{Introduction}
MHD modes in fusion plasma cover a large range of frequency spectrum, from several kHz low frequency tearing mode to the MHz high frequency whistle wave. However the sampling rate of the diagnostics such as Mirnov magnetic probes are limited in a particular experiment. When the sampling rate (or sampling frequency: $F_s$) of a signal is below the requirement ($F_s>2f_{sig}$) of Nyquist-Shannon sampling theorem, the measured frequency will not correctly remain the original frequency $f_{sig}$ value of the signal \cite{Shannon_1949_PIRE}. This error in measured frequency is called alias sampling (or under sampling). Alias sampling is common in our real life. For example, when a wheel is rotating fast enough, we can really see alias with our eyes as if the wheel is rotating backward. Because human eyes can only pick up images below 60 Hz. Which is to say, the sampling rate is Fs=60 Hz, when the wheel rotate higher than $F_s/2=30$ Hz, alias sampling occur to human eyes. The alias frequency formula has been well described in books about signal processing such as reference \cite{Bendat_2012_book, Press_1992_book}. In this research note, we will make a more detailed derivation of this relation. A now effect that alias sampling will not change the absolute structure of the MHD mode is also discovered. This discovery will help us to measure the correct high frequency mode structure using low frequency sampled diagnostics even when the Nyquist-Shannon sampling theorem is violated.

\section{Sampling of a sine function}
Since a time changing signal can be treated as a function $s(t)$ which can be Fourier decomposed as sine and cosine function series. If we can understand the alias sampling of a sine function, we can understand the alias effect on the whole frequency spectrum of the signal. The alias frequency originates from the periodicity of the sine function. The detailed mathematics goes like this. Say we have a sine function with frequency $\omega_1$ as: $sin(\omega_1 t+\phi_1)$. If we sampled this function at discrete time interval: $t_n = n\Delta$, n = 0, 1, 2, 3, ..., where $\Delta$ is the sampling time interval, related sampling rate is: $F_s = \frac{1}{\Delta}$. The continuous sine function under this sampling rate becomes discrete values as: $sin(\omega_1 t_n +\phi_1) = sin(\omega_1 n\Delta+\phi_1)$. We known the sine function has periodic relation as: $sin(x) = sin(x + 2k\pi)$, where k is an integer number. Thus we apply this relation to the sampled function and get the relation:

\begin{equation}
    \begin{array}{l l}
        sin(\omega_1 n\Delta+\phi_1) & = sin(\omega_1 n\Delta+\phi_1 + 2\pi kn)\\
        & = sin((\omega_1 + 2\pi k/\Delta)n\Delta + \phi_1)\\
        & = sin(2\pi(f_1 + kF_s)n\Delta + \phi_1)
    \end{array}
    \label{eq_alias_f_relation}
\end{equation}

Where $\omega_1 = 2\pi f_1$ is the angular velocity (or angular frequency). If we suppose that $f_1 = f_a$ is the signal we measure after discrete sampling and $f_0$ is the real frequency of the signal. Then any real signal with frequency $f_0$ will be sampled as alias frequency within the Nyquist frequency limit ($F_{Nyq} = F_s/2$) determined by sampling rate as $f_a\in [0,F_{Nyq}] = [0,F_s/2]$. Then the relation between alias frequency and real frequency is deduced from equation (\ref{eq_alias_f_relation}) as:

\begin{equation}
    \label{ramp_alias_freqeuncy}
    \begin{array}{l l}
        & f_0 = f_a+kF_s\\
        \Rightarrow & f_a = f_0 - kF_s
    \end{array}
\end{equation}

Since $f_a\in [0, F_s/2]$, we have $f_0 = f_a + kF_s \in [(k-1)F_s, (k-1/2)F_s]$, where $k = 1, 2, 3, ...$ . That is:

\begin{equation}
    f_a = f_0 - (k-1)F_s,\ ((k-1)F_s \leq f_0 \leq (k-1/2)F_s),\ k = 1, 2, 3, ...
\end{equation}

If we also consider the relation of sine function: $sin(x) = sin(\pi-x)$, we can get another relation:

\begin{equation}
    \begin{array}{l l}
        sin(\omega_1 n\Delta + \phi_1) & = sin(\pi - (\omega_1 n\Delta + \phi_1))\\
        & = sin(\pi - (\omega_1 n\Delta + \phi_1) + 2\pi kn)\\
        & = sin(2\pi(kF_s - f_1)n\Delta + \pi - \phi_1)
    \end{array}
\end{equation}

Like the derivation of equation (\ref{ramp_alias_freqeuncy}), we can get the relation between the alias and the real frequency under this condition as:

\begin{equation}
    \begin{array}{l l}
        & f_0 = kF_s - f_a\\
        \Rightarrow & f_a = kF_s - f_0
    \end{array}
\end{equation}

Similarly the range of $f_0$ is: $f_0 \in [(k-1/2)F_s,kF_s]$, where k = 1, 2, 3, ... . That is:

\begin{equation}
    f_a = kF_s - f_0,\ ((k-1/2)F_s \leq f_0 \leq kF_s),\ k = 1, 2, 3, ...
\end{equation}

Altogether, the relations between alias frequency and real frequency are:


\begin{equation}
    \label{alias_freqeuncy_relation}
    f_a =
    \begin{cases}
    f_0 - (k-1)F_s  &\quad ((k-1)F_s \leq f_0 \leq (k-1/2)F_s) \\
    kF_s-f_0  &\quad ((k - 1/2)F_s \leq f_0 \leq kF_s)
    \end{cases}
\end{equation}

Where k = 1, 2, 3, ... . is positive integer number. With help of equation (\ref{alias_freqeuncy_relation}), we can easily identify the alias frequency from spectrum.

\begin{figure*}
    \begin{minipage}{0.49\textwidth}
       \centering
	   \includegraphics[width=1.0\linewidth]{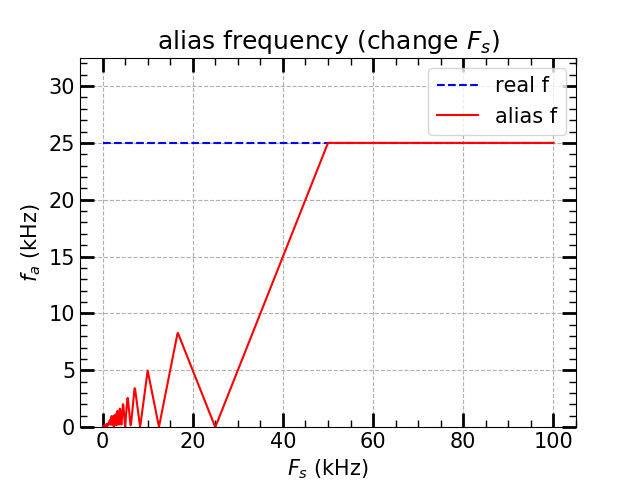}
	   \captionof{figure}{Alias frequency with constant signal frequency ($f_0 = $ 25 kHz) and changing sampling frequency ($F_s$ from 0 to 100 kHz).}
	   \label{fig_alias_freqeuncy_ramping_Fs}
    \end{minipage}%
    \hfill
    \begin{minipage}{0.49\textwidth}
	   \centering
	   \includegraphics[width=1.0\linewidth]{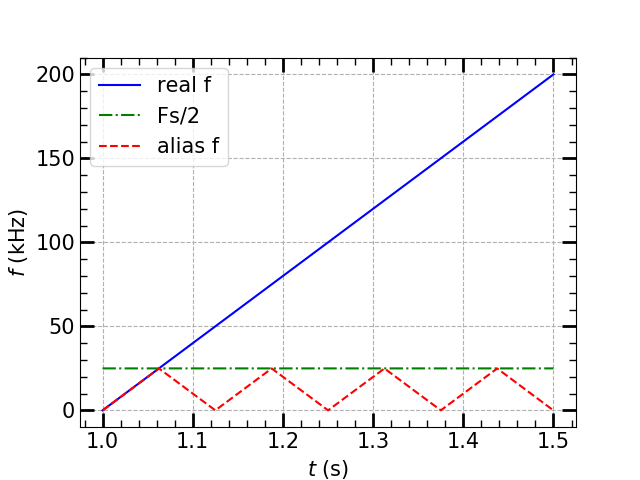}
	   \captionof{figure}{Alias frequency with constant sampling frequency ($F_s=$ 50 kHz) and ramping signal frequency ($f_0$ from 0 to 200 kHz).}
	   \label{fig_alias_freqeuncy_ramping_f0}
    \end{minipage}
\end{figure*}

Then let's see the alias frequency changing with sampling frequency. We generate  a test signal and hold $f_0=$ 25 kHz as constant, and ramping the sampling frequency $F_s$ from 0 to 100 kHz. Using this fake test signal as input, the simulation result is shown in Fig \ref{fig_alias_freqeuncy_ramping_Fs}.

From Fig \ref{fig_alias_freqeuncy_ramping_Fs} we can see the alias frequency is always smaller than the real signal frequency. When the sampling frequency is higher than $2f_0$, the sampled frequency is equal to the signal frequency. As the sampling frequency reduces below it, the alias frequency begins to ramp up and down many times and getting smaller and smaller.

Then we hold the sampling frequency as constant at $F_s=$ 50 kHz and ramp the signal frequency $f_0$ from 0 to 200 kHz, we can get this simulation result as Fig \ref{fig_alias_freqeuncy_ramping_f0} shows. The alias frequency ramps in zig-zag pattern within the range of Nyquist frequency band $[0, F_s/2]$.

\section{Alias effect on MHD mode number calculation}
The traditional Fourier transformation based spectrum calculation could only reveal the mode information in the frequency domain. The spectrogram (or Power Spectral Density/PSD) reveals knowledge of both mode spectrum as well as its evolution with time. To meet this requirement, a technique called Short Time Fourier Transformation (STFT) is invented to calculate the spectrogram of the diagnostics signal \cite{Allen_1977_IEEE,Griffin_1983_IEEE}.

\begin{figure*}
    \begin{minipage}{0.49\textwidth}
	   \centering
	   \includegraphics[width=1.0\linewidth]{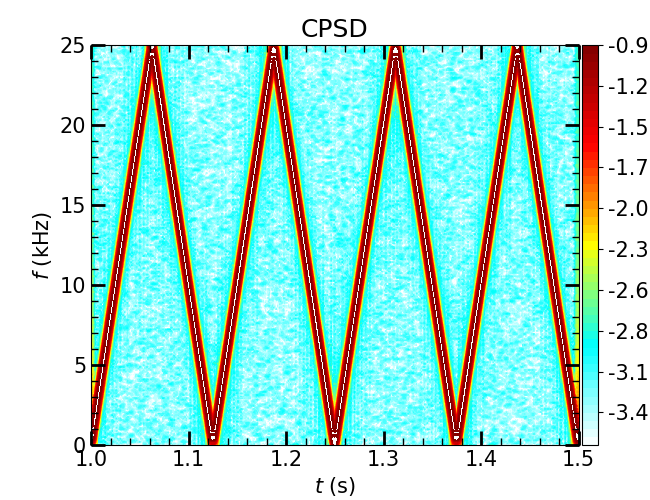}
	   \captionof{figure}{CPSD calculation from test n = 1 MHD mode with ramping frequency $f_0$ from 0 to 200 kHz at a sampling rate of $F_s = $ 50 kHz. The mode frequency for spectrogram calculation is set as same as Fig. \ref{fig_alias_freqeuncy_ramping_f0}.}
	   \label{fig_alias_CPSD}
    \end{minipage}%
    \hfill
    \begin{minipage}{0.49\textwidth}
	   \centering
	   \includegraphics[width=1.0\linewidth]{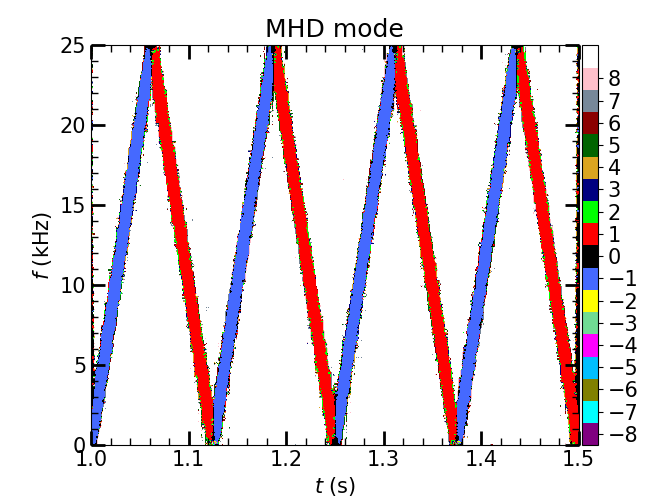}
	   \captionof{figure}{MHD mode number of test n = 1 MHD mode with ramping frequency from 0 to 200 kHz at $F_s = $ 50 kHz, where the white noise below a certain level is set as white color.}
	   \label{fig_alias_mode_number}
    \end{minipage}
\end{figure*}

To test alias effect on the MHD mode spectrogram, we create a fake $n=1$ mode signals with ramping frequency $f_0$ from 0 to 200 kHz by inserting a phase delay between to signals at $\Delta\phi=\pi/16$ and setting the fake coils geometry angle shift also at $\Delta \theta=\pi/16$. Then we set the sampling frequency as $F_s = 50 $ kHz and conduct the STFT transform to calculate its Cross Power Spectrum Density (CPSD). From the simulation result in Fig \ref{fig_alias_CPSD}, we can see the alias mode frequency ramping in the same pattern as Fig \ref{fig_alias_freqeuncy_ramping_f0} shows. Thus the alias sampling effect on the mode frequency is verified.

Then, let's see the effect of alias sampling on mode calculation. We take a first look on the calculation of toroidal mode number from adjacent Mirnov probes \cite{Hole_2009_RSI}. We note two adjacent Mirnov probes signal as: sig1(t) and sig2(t), their toroidal angle shift is $\Delta\theta$. If we conduct STFT to the two signals we can get the mode phases on the two probes as $\phi_1,\phi_2$. Then the phase shift between the two signals will be $\Delta\phi=\phi_2-\phi_1$, then the toroidal mode number is: $n=[\Delta\phi/\Delta\theta]$. The square bracket $[]$ here means to take the closest integer value.

Having known the process of mode number calculation, we can analytically discuss the alias effect on mode number. First, suppose the signals have the form: $sig1(t)=sin(\omega t+\phi_1), sig2(t)=sin(\omega t+\phi_2)$, then after alias sampling with $F_s=1/\Delta$, similar with the discussion in section 2, they become this:

\begin{enumerate}
	\item when $f_0=f_a+kF_s \in [(k-1)F_s,(k-1/2)F_s]$
	\begin{equation}
	    \begin{array}{l l}
     	    sig1(t_n)=sin(2\pi(f_a+kF_s)n\Delta+\phi_1)\\
     	    sig2(t_n)=sin(2\pi(f_a+kF_s)n\Delta+\phi_2)
	    \end{array}
	\end{equation}
	Thus the mode number becomes: $n_1=\frac{\phi_2-\phi_1}{\Delta\theta}=n_0$, which is same as the mode number with correct sampling.
	\item when $f_0=kF_s-f_a \in [(k-1/2)F_s,kF_s]$
	\begin{equation}
	\begin{array}{l l}
	sig1(t_n)=sin(2\pi(kF_s-f_a)n\Delta+\pi-\phi_1)\\
	sig2(t_n)=sin(2\pi(kF_s-f_a)n\Delta+\pi-\phi_2)
	\end{array}
	\end{equation}
	Thus the mode number becomes: $n=\frac{(\pi-\phi_2)-(\pi-\phi_1)}{\Delta\theta}=-\frac{\phi_2-\phi_1}{\Delta\theta}=-n_0$
\end{enumerate}

Altogether, we have the relation for mode number for alias sampling, suppose $n_0$ is the real toroidal mode number, and $n$ is the mode number we calculated after alias sampling. We have:


\begin{equation}
    n=
    \begin{cases}
        n_0 & (f_0=f_a+kF_s \in [(k-1)F_s,(k-1/2)F_s])\\
        -n_0 & (f_0=kF_s-f_a \in [(k-1/2)F_s,kF_s])\\
    \end{cases}
    \label{eq_alias_mode_fomula}
\end{equation}

The simulation result in Fig \ref{fig_alias_mode_number} shows the switching between $n = $1 and -1 modes in accordance with the zig-zag change of alias frequency in Fig \ref{fig_alias_CPSD}. Further simulation with n=-8,-7,...,3,2 modes also satisfy with our relation. Thus the simulation results agree perfectly with our alias mode number formula (\ref{eq_alias_mode_fomula}). Which is also to say the absolute mode number $|n|$ is not affected by the alias sampling. In another word, it means the alias sampling will not change the absolute mode structure of a MHD mode. Thus, from a low frequency alias sampled diagnostics signals we might also get useful information to estimate the correct mode structure of real high frequency MHD mode.

\section{Alias spectrum amplitude}
Another concern is the spectrum amplitude of the MHD mode under alias sampling. {Will the mode strength becomes weak as alias appear under a low sampling rate?} Suppose a MHD mode has frequency $f_0$ and amplitude $A_0$, its alias harmonics shall be $kF_s \pm f_0$, the related alias amplitudes are: $A_{ak\pm},\ k=1, 2, 3,...$. Then after discrete alias sampling, all the alias harmonic frequencies will be concentrated within the Nyquist band ($f_a \in [0, Fs/2]$). Since waves propagation and Fourier transformation all satisfy the principle of linear superposition, the concentrated alias amplitude within Nyquist band should also be a linear summation of all the alias harmonic amplitudes as:

\begin{equation}
    A_a = A_0 + A_{a1-} + A_{a1+} + A_{a2-}+ A_{a2+} + ...= A_0 + \sum_k (A_{ak-} + A_{ak+})
    \label{eq_alias_amplitude}
\end{equation}

\begin{figure*}
    \begin{minipage}{0.49\textwidth}
	   \centering
	   \includegraphics[width=1.0\linewidth]{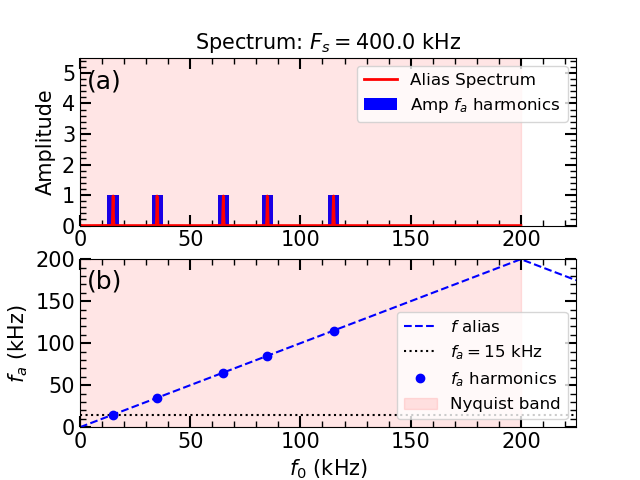}
	   \captionof{figure}{Real spectrum of $f_0$ harmonics with the Nyquist frequency band with proper sampling rate at $F_s=400$ kHz. (a) Full specturm and real harmonics amplitude. (b) Alias frequecy of $f_0$ harmonics.}
	   \label{fig_test_full_spectrum}
    \end{minipage}%
    \hfill
    \begin{minipage}{0.49\textwidth}
	   \centering
	   \includegraphics[width=1.0\linewidth]{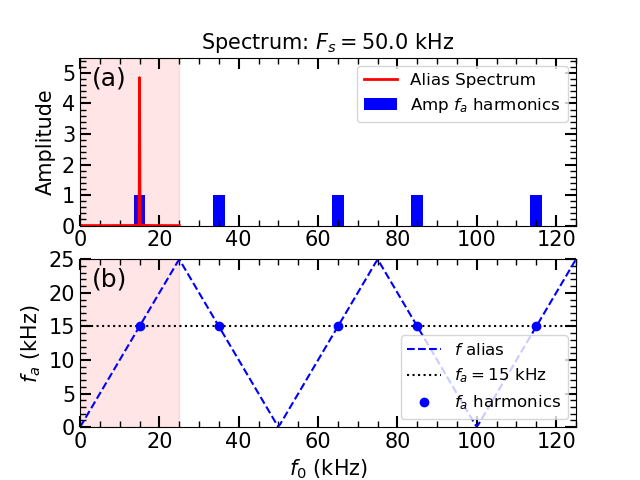}
	   \captionof{figure}{Alias spectrum of $f_0$ harmonics beyond Nyquist frequency band with alias sampling rate at $F_s=50$ kHz. (a) Alias specturm and real harmonics amplitude. (b) Alias frequecy of $f_0$ harmonics.}
	   \label{fig_test_alias_spectrum}
    \end{minipage}
\end{figure*}

The relation has been tested with simulation results. Suppose we have 3 alias harmonics modes in a signal at $f_0=15$ kHz, $f_{a1-}=50-15$ kHz, $f_{a1+}=15+50$ kHz, $f_{a2-}=50*2-15$ kHz, $f_{a2+}=15+50*2$ kHz, each with equal amplitude $A_0=A_{a1-}=A_{a1+}=A_{a2-}=A_{a2+}=1$. The frequencies are so set that they will become alias harmonics when sampled at $F_s=50$ kHz. Under adequate sampling rate at: $F_s=400$ kHz, the spectrum amplitude calculation in Fig. \ref{fig_test_full_spectrum} (a) shows the same result as the setting values. However, if we reduce the sampling rate to: $F_s=50$ kHz, which no long satisfy the Nyquist-Shannon sampling theorem, the calculated alias harmonics in spectrum are not able to be distinguished from each other as Fig. \ref{fig_test_alias_spectrum} (a) shows. And their summed up amplitudes are found to be equal with the calculated amplitude in spectrum as: $A_a=5=A_0+A_{a1-}+A_{a1+}+A_{a2-}+A_{a2+}$. Thus the simulation result proves our formula (\ref{eq_alias_amplitude}) that under alias sampling, the alias amplitude will be a summation of all the alias harmonics.

In real experimental condition, there usually has one single MHD frequency $f_0$. Then under alias sampling, it will only produce one alias frequency within Nyquist band. All the other alias harmonics amplitudes are close to zero at white noise level. Which is to say, the amplitude of this alias frequency will be close or equal to the real MHD mode amplitude:


\begin{equation}
    A_a = A_0 + noise \approx A_0
\end{equation}

\section{Summary}
The alias frequency formula is derived in detail with an example of discrete sampling of a sine wave. New discussion of the alias sampling effect on the calculation of MHD mode number shows the alias sampling will only change the sign of the MHD mode number. It will not change the absolute structure of the MHD mode. And the spectrum amplitude of the alias mode will be close or equal to the original high frequency mode. Thus using low frequency alias sampled diagnostics signals to estimate the real high frequency MHD mode structure is also possible.


\providecommand{\noopsort}[1]{}\providecommand{\singleletter}[1]{#1}%

\bibliographystyle{unsrt}
\end{document}